\documentclass[]{aa} 
\usepackage{graphicx}
\usepackage{txfonts}
\usepackage{amssymb}
\usepackage{lineno}
\usepackage{txfonts}
\usepackage{natbib}   
\usepackage{amstext}
\usepackage{mathabx}
\usepackage{multirow}
\usepackage{ulem}
\usepackage{color}
\usepackage{pifont}
\usepackage{lscape}
\usepackage{mathtools}
\usepackage{enumitem}

\usepackage{hyperref}
\begin{document} 


	\title{Multi-antenna probing of absorbing regions inside and outside Cassiopeia A}
	\titlerunning{Multi-antenna probing of absorbing regions}

   \author{Lev~A.~Stanislavsky\inst{1}
          \and
          Ihor~N.~Bubnov\inst{1}
          \and
					Aleksander~A.~Stanislavsky\inst{1}
					\and
					Philippe~Zarka\inst{2}
					\and
					Alan~Loh\inst{2} 
					\and
					C\'edric~Viou\inst{3}
					\and
					Aleksander~A.~Konovalenko\inst{1}
					\and
					Anatolii~I.~Brazhenko\inst{4}
					\and
          Anatolii~V.~Frantsuzenko\inst{4}
          }

   \institute{Institute of Radio Astronomy, National Academy of Sciences of Ukraine, Mystetstv St., 4, 61002 Kharkiv, Ukraine\\
              \email{lev.stanislavskiy@gmail.com; a.a.stanislavsky@rian.kharkov.ua}
					\and
          LESIA, Observatoire de Paris, CNRS, Universit\'e PSL, Sorbonne Universit\'e, Universit\'e Paris Cit\'e, CNRS, Place J. Janssen, 92190 Meudon, France
          \and
					Observatoire Radioastronomique de Nan\c{c}ay (ORN), Obs. Paris, CNRS, PSL, UO, Nan\c{c}ay, France
          \and
					Poltava Gravimetric Observatory, Subbotin Institute of Geophysics, National Academy of Sciences of Ukraine, Myasoyedov St., 27/29, 36029 Poltava, Ukraine
             }

   \date{Received ; Accepted }

  \abstract
   {Cassiopeia A occupies an important place among supernova remnants (SNRs) in low-frequency radio astronomy. Located in our Galaxy, this powerful radio source emits synchrotron radiation that propagates through the SNR environment and the ionized interstellar medium. The analysis of its continuum spectrum from low frequency observations reveals the evolution of the SNR absorption properties over time and suggests a method for probing unshocked ejecta and the SNR interaction with the circumstellar medium (CSM).}
   {In this paper we present low-frequency measurements of the integrated spectrum of Cassiopeia A to find the typical values of free-free absorption parameters towards this SNR in the middle of 2023. We also add new results to track its slowly evolving and decreasing integrated flux density.}
   {We used the New Extension in Nan\c{c}ay Upgrading LOFAR (NenuFAR) and the Ukrainian Radio Interferometer of NASU (URAN--2, Poltava) for measuring the continuum spectrum of Cassiopeia A within the frequency range of 8--66 MHz. The radio flux density of Cassiopeia A relative to the calibration source, the radio galaxy Cygnus A, has been obtained on June--July, 2023 with two sub-arrays for each radio telescope, used as a two-element correlation interferometer.}
   {We measured magnitudes of emission measure, electron temperature and an average number of charges of the ions for both internal and external absorbing ionized gas towards Cassiopeia A from its integrated spectrum. Generally, their values are comparable to those presented by \citet{Stanislavsky2023}, but their slight changes show the evolution of free-free absorption parameters in this SNR. Based on high accuracy of the measurements, we have detected the SNR--CSM interaction. This led to the fact that the maximum of this continuous spectrum, decreasing in intensity, changed in frequency from past higher values to lower ones over time. Probably, such changes occur periodically in which the spectral peak can move in the direction of both higher frequencies and lower ones. This effect is mainly caused by the evolution of emission measure outside Cassiopeia A.}
   {The integrated flux-density spectrum of Cassiopeia A obtained with the NenuFAR and URAN--2 interferometric observations opens up new possibilities for continuous monitoring the ionized gas properties in and around Cassiopeia A to observe the evolution of unshocked ejecta and the SNR-CSM interaction in future studies.}

\keywords{ISM: individual objects: Cassiopeia A -- galaxies: individual: Cygnus A -- methods: observational -- instrumentation: miscellaneous -- telescopes}

   \maketitle

\section{Introduction}
The radio emission of Cassiopeia A (Cas A; 3C461, G111.7-2.1) is of a synchrotron nature, the important role of which was first pointed out by I.~\citet{Shklovsky1954}. Like other supernova remnants (SNRs), Cas A emits radio waves due to shocks accelerating the relativistic electrons. For a long time it has been the strongest extra-solar radio source used as a calibrator for different radio astronomy instruments \citep{Baars1977}, but due to adiabatic expansion the young core-collapse remnant exhibits a secular decrease in flux density \citep{Patnaude2011,Vinyaikin2014,Trotter2017}. This effect has relegated it to the second place, swapping with Cygnus A (Cyg A; 3C405). Moreover, the flux-density decline rate slows down over the years. Therefore, the continuous monitoring of this radio source is important not only for calibration purposes, but the radio observations give unique information on the evolution of Cas A and its interaction with the surround environment \citep{Kassim1995,Gasperin2020}. The radio spectrum of SNRs is very sensitive to absorption processes at low frequencies (< 100 MHz). The absorption is caused by ionized gases in and around SNRs, and they have various values of free-free absorption parameters evolving over time. Also Cas A is interesting in that it is one of the few SNRs, clearly exhibiting unshocked ejecta in low-frequency observations \citep{Raymond2018}. Using the morphology and spectrum of Cas A obtained with the Low Frequency Array (LOFAR, \citealp{vanHaarlem13}) at low frequencies, the mass and density in the unshocked ejecta were found by \citet[and references therein]{Arias2018}. On the other hand, the external absorption in the interstellar medium (ISM) also influences on the radio spectrum of Cas A. Therefore, both types of absorption are taken into account for fitting the experimental radio spectra of the SNR to theoretical models. Confirming the results, the independent study of the integrated radio spectrum of Cas A in the epoch of 2019 was implemented by \citet{Stanislavsky2023}. These low-frequency observations used a two-element correlation interferometer from two sub-arrays of the Giant Ukrainian radio telescope (GURT) each of which had 5 $\times$ 5 active crossed dipoles. This approach implements the relative method for measuring the radio emission flux density of Cas A with respect to Cyg A, which serves as an absolute standard calibrator \citep{Braude1962,Perley2017}. The latter radio source is a nearby ($z\approx0.056$) bright radio galaxy lying in the Galactic plane and has been intensively studied for as long as Cas A. It is widely accepted that Cyg A cannot possibly be variable on ``human-lifetimes'' at low frequencies, since only its sub-parsec component (the nucleus) capable of varying on suitable timescales constitutes less than 0.1\% of the total flux density at the frequencies. \citet{Ryle1949} were the first to make absolute measurements of Cas A against the flux density of Cyg A at 81.5 MHz. This method has proven to be very effective \citep{Erickson1975}. If Cas A has a secular decline in flux density, Cyg A does not, which makes it a good calibrator for measuring the flux density. Its accuracy is about 3--5\% at low frequencies. All this allows for the required measurements to be taken with the highest level of accuracy.

In this paper we report the results of radio observations carried out in June--July, 2023. The aim is to measure the  flux density of Cas A in the epoch of 2023.5 to detect the evolution of free-free absorption parameters in and around this SNR with respect to the data of previous GURT measurements made in 2019 \citep{Stanislavsky2023}. Due to the Russian invasion and occupation in February 2022, the GURT radio telescope was disabled. New observations of Cas A at 8--66 MHz have been fulfilled with other radio telescopes. From their sub-arrays we organized a two-element correlation interferometer for each radio telescope. Their accuracy is the same as in the GURT observations at 16--72 MHz. By means of the measured flux-density spectrum of Cas A, we determine typical values of emission measure, temperature and an average number of charges of the ions in absorbing regions, and to estimate the secular decline of the radio flux emitted by Cas A in the middle of 2023, as well as its spectral index. Finally, we compare our results with previous ones and discuss their differences with possible reasons of these effects. 

\section{Observation method}

\subsection{NenuFAR interferometer}\label{section2_1}
New extension in Nançay upgrading LOFAR (NenuFAR) is a new radio telescope constructed and commissioned at the Nan\c{c}ay Radioastronomy Observatory \citep{Zarka2020,Girard2023}. It can provide observations at 10--85 MHz in different modes such as a standalone beamformer, a standalone imager, a waveform snapshots recorder and, in the near future, a giant low-frequency station of the LOFAR array. In our observations we used the simplest mode, a standalone beamformer, but with some modifications. Their description is given below.

\begin{figure}
\centerline{\includegraphics[width=\columnwidth]{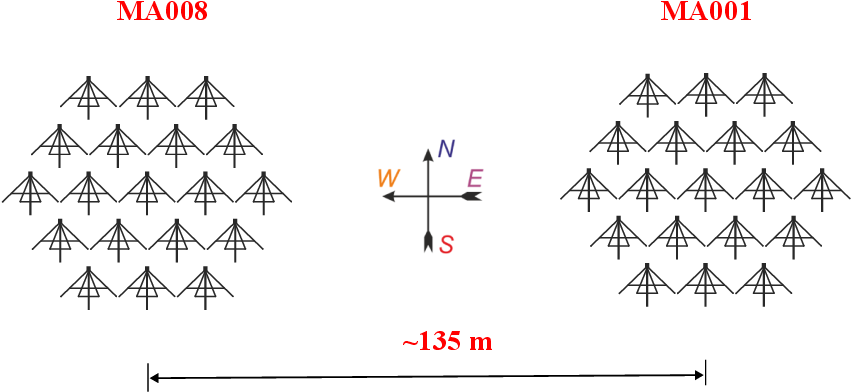}}
\caption{\label{fig1_nenufar}
Configuration of NenuFAR mini-arrays used in the radio observations of Cas A and Cyg A.}
\end{figure}

The NenuFAR radio telescope includes 96 mini-arrays, each of which have 19 elementary antennas arranged in a hexagonal shape. Each elementary antenna is built from LWA-style crossed inverted-V dipoles oriented at 45$^\circ$ from the meridian like LOFAR dipoles. The geographical latitude of NenuFAR (47.38$^\circ$N) allows for Cas A and Cyg A to be observed in close zenith angles at the time of their upper culminations. The observations were carried out with two mini-arrays, MA001 and MA008, forming a baseline of 92~m along west-east (Figure~\ref{fig1_nenufar}). These mini-arrays have identically oriented crossed dipoles. The use of a short-spacing two-element interferometer in itself has certain advantages. This is simplicity, a sufficient signal-to-noise ratio of the chosen configuration, and the ability to find integrated radio spectra from these sources. This method is very similar to the one used by \citet{Ryle1952}. If the 2 $\times$ 2 polarizations analog signals with MA001 and MA008 are sent to the main NenuFAR standalone receiver, then only an additive interferometer is implemented, which does not exclude the intense contribution of the Galactic background radiation at low frequencies. This makes it very difficult to compare the flux density received from different sources. Therefore, we applied the following procedure to overcome this problem. The signals from MA001 and MA008 were split into two equal parts from each mini-array. This allowed us to record three different signals ($A^2$, $B^2$, and $(A+B)^2$) from which the correlation response $AB$ is obtained by simple algebraic operations during the data processing. The best result was recorded in the observations of June 24, 2023.

Using the python package nenupy\footnote{https://nenupy.readthedocs.io/en/latest/index.html}, it is straightforward to calculate the effective area of such a mini-array in the direction of zenith. At 40 MHz the value equals to about 356 m$^2$. We recall that the same effective area is also typical for each GURT sub-array of 5 $\times$ 5 crossed dipoles at the same frequency. If $\Delta\nu$ = 200~kHz and $\Delta\tau$ = 0.5~sec are the frequency and the time resolutions, then relative sensitivity of radio-source measurements is about $10^{-2}$. According to \citet{Stanislavsky2023}, this sensitivity, related to Cas A and Cyg A, is quite sufficient to produce highly reliable data of good quality. 

\subsection{URAN--2 interferometer}\label{section2_2}
\begin{figure}
\centerline{\includegraphics[width=\columnwidth]{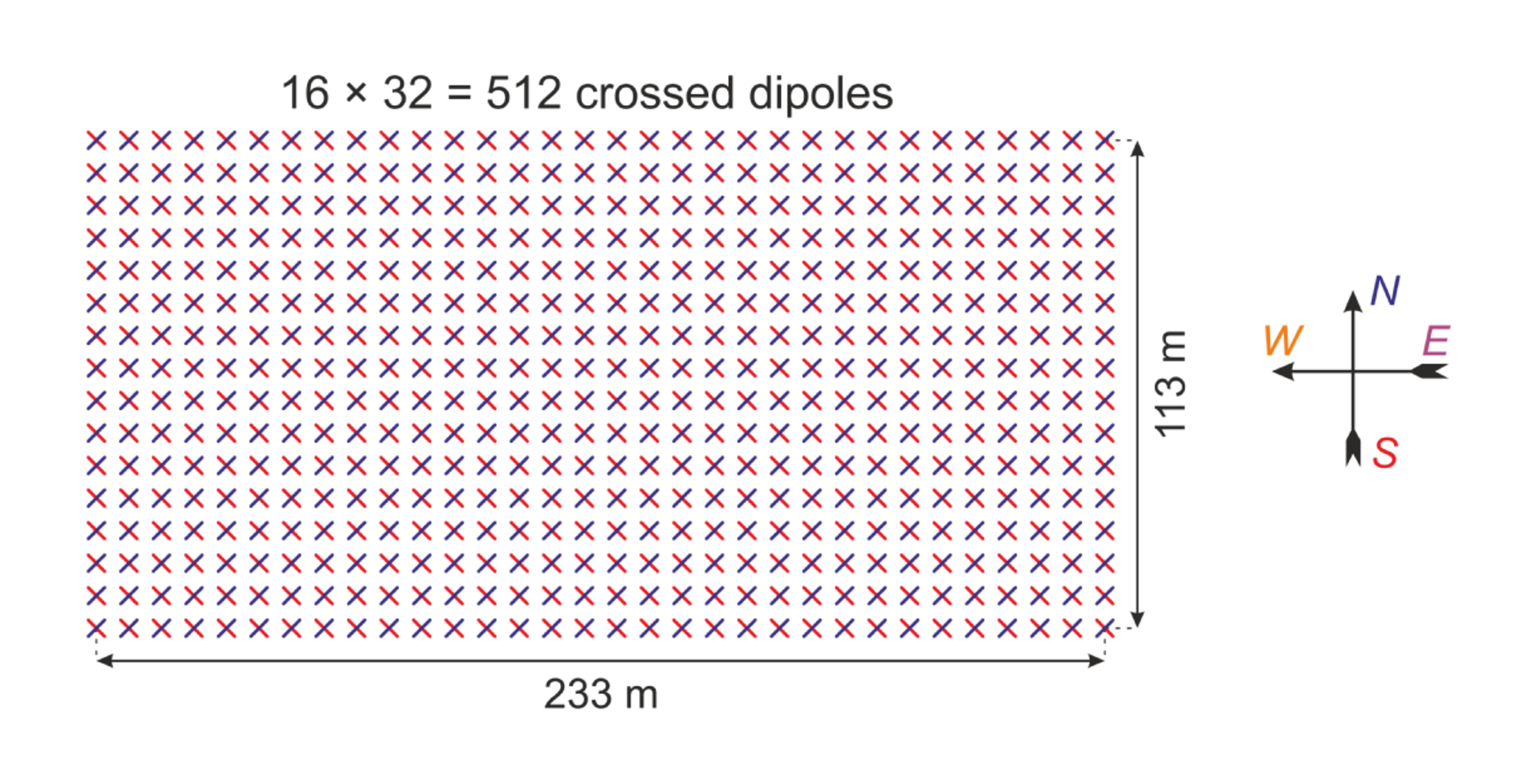}}
\caption{\label{fig2_base}
Location of URAN--2 crossed dipoles shown by crosses. The interferometer consists of two equal halves of this antenna. Each has 16$\times$16 dipoles.}
\end{figure}

URAN--2 is one of four famous Ukrainian Radio Interferometers under the auspices of the National Academy of Sciences of Ukraine located in Poltava \citep{Brazhenko2005}. Its geographical latitude is close to 49.6$^\circ$N and very convenient for the radio observations of Cas A and Cyg A sources having the same zenith angles at their upper culminations. The URAN--2 radio telescope is intended for receiving radio emission within the frequency range of 10--30 MHz \citep{Konovalenko2016}. This is a rectangular arrays of 16 $\times$ 32 crossed dipoles. However, its observational potential can be extended for Cas A and Cyg A. For such powerful signals coming from the radio sources, their observation at 8--40 MHz is quite feasible. They were carried out with two equal halves (sub-arrays) of the URAN--2 full antenna each of which included 16 $\times$ 16 crossed dipoles (Figure~\ref{fig2_base}). The effective area of each URAN--2 half is about three times larger than any NenuFAR mini-array at 40 MHz in its entirety. Therefore, the URAN--2 sub-array exhibits a higher sensitivity. The two-element interferometer based on the URAN--2 has a baseline along west-east. The distance between their phase centers is 120 m. The interferometric base is longer in two times than one of the GURT interferometer in 2019 \citep{Stanislavsky2023} and comparable in length to the base of NenuFAR used for this experiment. Observations with URAN--2 were recorded using the ADR (advanced digital receiver) which is a standard backend for the GURT observations \citep{Zakharenko2016} and was available for the URAN--2 observations.

\subsection{Observations}

Our observations were divided into several sessions. The first of them occurred on June 24, 2023 with NenuFAR. The next observations lasted from June 26 to July 5, 2023 with URAN--2. Later the second session with NenuFAR mini-arrays was implemented at night from July 11-12, 2023. Each seance of observations was accompanied by records of radio emission from Cas A and Cyg A separately one after another in the transit mode. To measure the flux-density measurement of Cas A, we used the Cyg A as a calibration source with the well-known flux density which can be considered practically unchanged in comparison with Cas A. The ratio of radiation flux densities of the Cas A and Cyg A sources is related to the ratio of the measured signal powers in the culmination direction of these sources (see detailed discussions of this method in \citealp{Stanislavsky2023}).

\begin{figure}
\centerline{\includegraphics[width=\columnwidth]{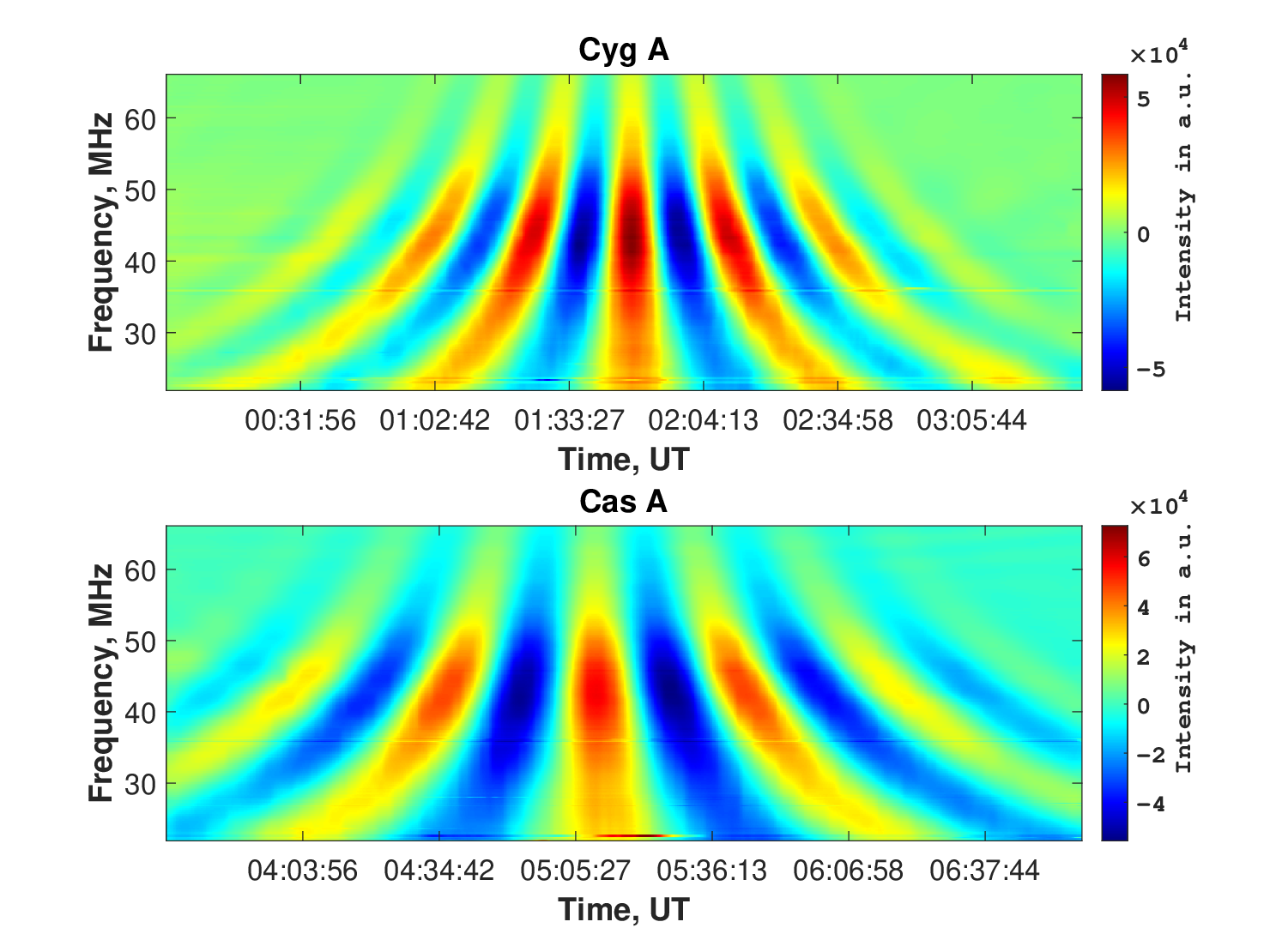}}
\caption{\label{fig1_resp}
Interferometric responses obtained from the observations on June 24, 2023 with two mini-arrays of NenuFAR. The color-coded intensity decline of this figure towards lower frequencies is explained by the amplitude-frequency response of this antenna system that includes preamplifiers, filters and others. But this does not affect the spectral results, since we use a ratio of measurements for Cas A and Cyg A.}
\end{figure}

At first we planned to make observations with NenuFAR mini-arrays, starting with 16 MHz just like the GURT observations in 2019 for their comparison. The NenuFAR beamformer can manage 768 beamlets (of 195.3125 kHz = 200 MHz/1024) at a time, representing a total frequency band of 150 MHz. As we distribute these 768 beamlets in 3 beams ($A^2$, $B^2$ and $(A+B)^2$), we can cover 150/3 = 50 MHz for each beam. Hence the 16--66 MHz band observed. This frequency boundary is slightly smaller than 72 MHz for the GURT observations in 2019. Special filters of NenuFAR cut off these recordings below 15 MHz. Dynamic spectra in the {\it DynSpec data} format were obtained with the coarsest possible time resolution 84 ms under the frequency resolution of 12.2 kHz. However, although the NenuFAR observations were made at night, intensive radio frequency interference (RFI) greatly hindered the emergence interferometric responses below 21 MHz. This situation was on June 24. For that day the interferometric responses of Cas A and Cyg A are presented in Figure~\ref{fig1_resp}. In the repeated observations on July 11-12, the records were damaged by RFI below 23 MHz, but at 23--66 MHz the ratio of interferometric responses from Cas A and Cyg A for both NenuFAR observation seances were very close over the entire frequency range. To expand the frequency range of observations and obtain data up to the spectral maximum of Cas A at low frequencies, we made additional observations with another telescope, namely with the URAN--2.

Radio records of the URAN--2 observations with the ADR had the time resolution in 1 s under the frequency resolution of 9.766 kHz. As each ADR permits the measurement of cross spectra, one entry of the ADR was connected to one half of the URAN--2 array, and the second entry received signals from another half. The measurements of the flux-density ratio of Cas A and Cyg A were made between 8 MHz to 40 MHz. Unfortunately, at $<$ 20 MHz the URAN--2 records of the radio sources were damaged by intensive RFI in the observations at June 26-30, 2023. Therefore, we did not use them in the data processing. The best records with URAN--2 were obtained on July 3, 2023. Therefore, our data processing mainly covers those data. Eventually, the average measurement errors for NenuPAR and URAN-2 are similar to the errors of the GURT  observations implemented in 2019.

\begin{figure}
\centerline{\includegraphics[width=\columnwidth]{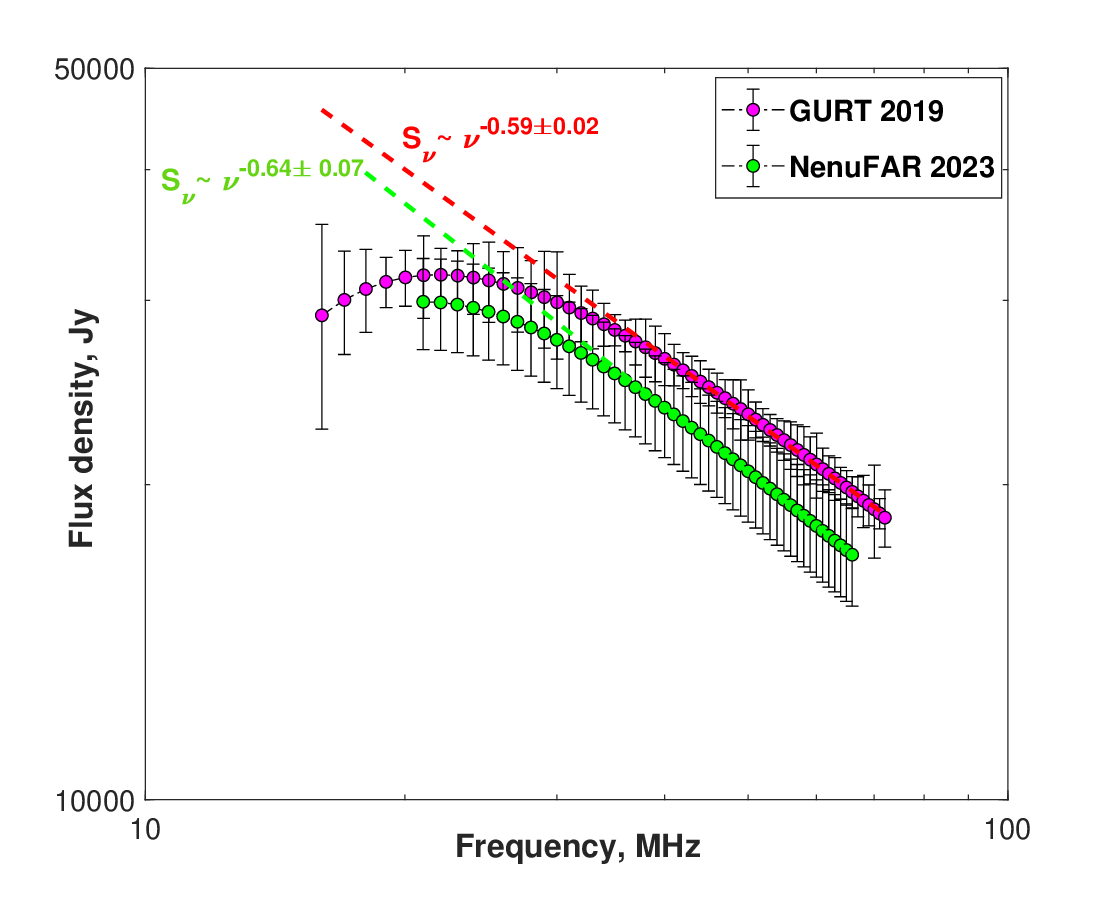}}
\caption{\label{fig5_CasA_flux}
Radio spectrum of Cas A obtained by two mini-arrays of NenuFAR in the range of 21--66 MHz. Red and green dash lines indicate spectral indexes of the measured radiation at 38--66 MHz for Cas A obtained with the GURT \citep{Stanislavsky2023} and the NenuFAR  radio telescopes, respectively.}
\end{figure}

\section{Data analysis}

\subsection{Spectral measurements with NenuFAR and URAN--2}

Our data analysis covers mainly the frequency range 8--66 MHz. The Cas A flux-density spectrum at low frequencies was concatenated from two parts. From 8 MHz to 21 MHz, we used radio records from URAN--2, with NenuFAR data added in the frequency range of 21--66 MHz. It should be noticed that the spectrum of Cas A with URAN--2 was also observed from 21 MHz to 40 MHz, and it demonstrated a good agreement with the spectral results with NenuFAR. The largest difference between them did not exceed 2.5\% below 30 MHz, and at lower frequencies it was fractions of a percent. Both instruments gave similar errors in measurements. The flux density of Cas A on a selected frequency grid with the step 1 MHz from the NenuFAR observations is shown in Figure~\ref{fig5_CasA_flux}. For comparison the spectrum from the observation in the middle of 2023 is presented with the spectral results obtained by the GURT sub-arrays in 2019. The intensity of the radio emission from Cas A has clearly decreased over the past 4.5 years. Below, we will discuss other features of the resulting spectrum.

\subsection{Model of free-free absorption}\label{model}

Here a brief review about the model of the radiation source inside Cas A and its impact on the environment around it will come in handy. As usual, we start with the description of a homogeneous cylindrical synchrotron source from the book of \citet{Pacholczyk1970}. This representation has the form
\begin{equation}
J_\gamma(z)\sim z^{2.5}\left(1-\exp\left(-z^{-2-\gamma/2}\right)\right)\,,\label{eq5}
\end{equation}
depending from the normalized frequency $z = \nu/\nu_1$ as well as from $\gamma$ characterizing a power-law distribution of electrons as $N(E)=N_0E^{-\gamma}$. Consequently, the ``bare'' radio source (without taking ionized gases from ejections into account) emits the flux density in the form of $S_\nu=AJ_\gamma(\nu/\nu_1)$ which includes three parameters: $\gamma$, $\nu_1$ and $A$. They describe the synchrotron emission from the SNR, whose structure and features are roughly known.

Since this synchrotron radiation propagates in reality through the unshocked ejecta of this SNR and the surround medium, where it is absorbed by intervening ionized gas, the absorption is determined by two components, internal and external \citep{Kassim1995,DeLaney2014,Arias2018}. The internal part occurs due to the unshocked ejecta mentioned above. The external component comes from the ISM around Cas A, where the remnant shock front has not yet reached. Eventually, the flux density is expressed as
\begin{equation}
S_\nu = \left(S_{\nu,\rm front}+S_{\nu,\rm back}e^{-\tau_{\nu,\rm int}}\right)e^{-\tau_{\nu,\rm ISM}}\,,\label{eq6}
\end{equation} 
where the free-free optical depths $\tau_{\nu,\rm int}$ and $\tau_{\nu,\rm ISM}$ determine the contribution of internal and external absorption, respectively. inside Cas A the relative synchrotron brightness is dependent on the front and back halves of the remnant differently. It can be taken into account using an additional parameter. Then the absorption on the two halves of the shell is considered as $S_{\nu,\rm front}=f_aS_\nu$ and $S_{\nu,\rm back}=(1-f_a)S_\nu$. The free-free optical depth follows the Rayleigh–Jeans approximation (see more details in the well-known book of \citealp{Wilson2009}). Thus, the optical depth value is described as
\begin{equation}
\tau_\nu = 3.014\times 10^4\,Z\left(\frac{T}{\rm K}\right)^{-3/2}\left(\frac{\nu}{\rm MHz}\right)^{-2}\left(\frac{EM}{\rm pc\ cm^{-6}}\right)g_{\rm ff}\,.\label{eq7}
\end{equation}
This value is determined by three parameters: emission measure, $EM$, average number of ion charges, $Z$, and the electron temperature of the absorbing medium, $T$. The formula in Eq. (\ref{eq7}) also includes the Gaunt factor $g_{\rm ff}$, but it can only depend on $Z$ and $T$, namely
\begin{equation}
g_{\rm ff}=\left\{
\begin{array}{ll}
\ln\left[49.55\,Z^{-1}\left(\frac{\nu}{\rm MHz}\right)^{-1}\right]+1.5\ln\frac{T}{\rm K} &  \\
1 \qquad \textrm{for}\qquad\frac{\nu}{\rm MHz}\gg\left(\frac{T}{\rm K}\right)^{3/2}\,.  & 
\end{array}\right.\label{eq8}
\end{equation}
The emission measure along the line-of-sight is of the following integral
\begin{equation}
\frac{EM}{\rm pc\ cm^{-6}}=\int_0^\frac{s}{\rm pc}\left(\frac{n_e}{\rm cm^{-3}}\right)\,\left(\frac{n_i}{\rm cm^{-3}}\right)\,d\left(\frac{s}{\rm pc}\right)\,,\label{eq9}
\end{equation}
where $n_e$ and $n_i$ are the electron density and the ion density, respectively. Summing up, it should be pointed out that the complete model of this radio source with free-free absorption depends on ten parameters, which need to be drawn from the spectral measurements.

\begin{figure}
\centering
\includegraphics[width=\columnwidth]{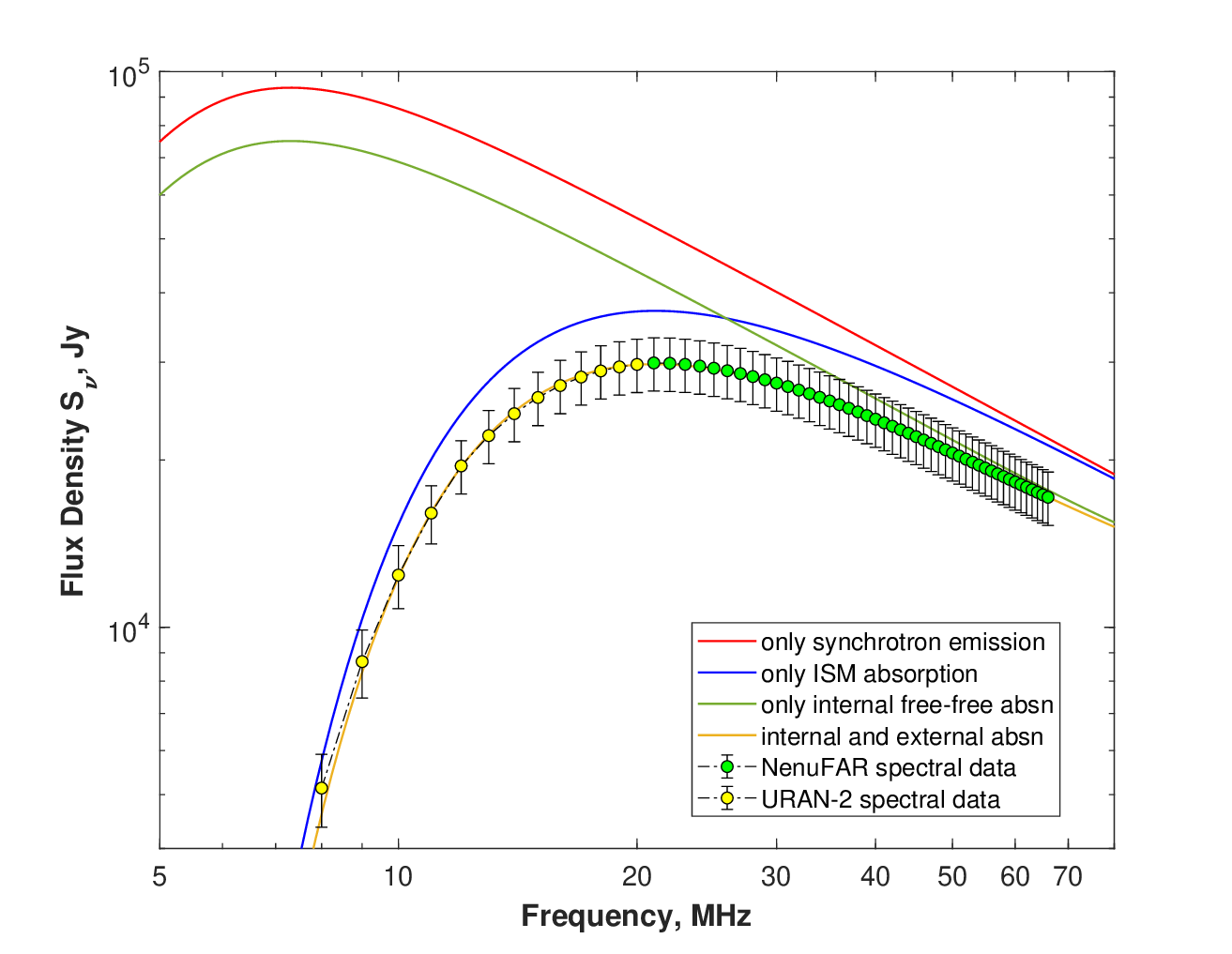}
\caption{\label{fig7_Comp} Impact of thermal absorption on the radio spectrum of Cas A, according to the spectral measurements at 8--66 MHz with NenuFAR and URAN--2 in the middle of 2023. Fitting of free-free absorption parameters in the model of Eqs.~(\ref{eq5})-(\ref{eq8}) indicates the contribution of different forms of absorption.}
\end{figure}

\subsection{Absorption evolution towards Cas A}

The flux-density spectrum of Cas A from the GURT observations made it possible to determine the values of absorption parameters in the epoch of 2019 \citep{Stanislavsky2023}. The same procedure can be applied to the spectral results shown in Figures~\ref{fig7_Comp} from the subsequent measurements with NenuFAR and URAN-2 in the middle of 2023. To fit our experimental spectrum to the theoretical curve of $S_\nu$ in Eq. (\ref{eq6}), we need to set up the model parameters. This goal is achieved through the nonlinear least-squares fitting. As initial values in this procedure, we use the results obtained for the GURT spectrum of Cas A from observations of 2019. Consequently, the obtained values of free-free absorption parameters are collected in Table~\ref{tab1}. They show close similarities with previous results of 2019. Their errors listed in Table~\ref{tab1} were derived from errors of the measured spectrum. The comparison between theoretical and experimental results is represented in Figure~\ref{fig7_Comp}. In particular, the unabsorbed synchrotron spectrum of Cas A is characterized by the following values: $\gamma\approx2.54$, $\nu_1\approx5.97$~MHz, and $A\approx140000$~Jy, where we recall 1 Jy = 10$^{-26}$ W m$^{-2}$ Hz$^{-1}$. Comparing with the results of 2018 \citep{Arias2018}, which had $\gamma\approx2.52$, $\nu_1\approx6.15$~MHz, and $A\approx140500$~Jy, the synchrotron source of 2023.5 has undergone insignificant changes.

\begin{table*}
\caption{Parameters of absorbing regions in Cas A obtained from the integrated continuum flux-density spectrum measured with the NenuFAR and URAN--2 interferometers.}%
\label{tab1}
\centering
\small
\resizebox{\linewidth}{!}{
\begin{tabular}{cccccccc}
\hline\hline
 &  &  &  &  & &  & \\
\textbf{Parameters} & $EM_{\rm int}$ (pc cm$^{-6}$)  & $EM_{\rm ISM}$ (pc cm$^{-6}$) & $f_a$ & $T_{\rm int}$ (K) & $Z_{\rm int}$ & $T_{\rm ISM}$ (K) & $Z_{\rm ISM}$ \\
 &  &  &  &  & &  & \\
\hline
 & & & & & & & \\
\textbf{Values} & 37.4 $\pm ^{0.02} _{0.01}$ & 0.15 $\pm ^{0.07} _{0.01}$ & 0.80 $\pm ^{0.08} _{0.06}$ & 100 $\pm ^{0.01} _{0.01}$ & 2.50 $\pm ^{0.09} _{0.1}$ & 20 $\pm ^{0.01} _{0.01}$ & 0.50 $\pm ^{0.01} _{0.13}$ \\
 & & & & & & & \\
\hline
\end{tabular}
}
\end{table*}

The important feature the experimental flux-density spectrum of Cas A is a peak location in frequency and its magnitude. Following the GURT observations in 2019, the spectral maximum reached 32190 $\pm$ 1894~Jy at 22.345 $\pm$ 0.5 MHz, using the corresponding fitted spectrum of this radio source. As applied to the observations with NenuFAR and URAN-2, the peak is located near 21 MHz. Therefore, it is difficult to estimate its properties only from the NenuFAR data at the edge of their frequency range, but here comes to the aid of observations with URAN--2 at 8--40 MHz. From the URAN--2 data the spectral peak of Cas A tends to 29880 $\pm$ 2980~Jy at 21.5 $\pm$ 0.2 MHz in the middle of 2023. The spectral index in 2019 was $\alpha=0.59\pm 0.02$ at 38--70 MHz, and the NenuFAR observations of 2023.5 give a similar value, namely, $\alpha=0.64\pm 0.07$, at 38--66 MHz. Both results are shown in Figure~\ref{fig5_CasA_flux}. The frequency 38 MHz is often used to compare secular decline changes in the ratio of flux densities for the radio emission from Cas A and Cyg A. In 2019 the value was 1.068 $\pm$ 0.068, while later in 2023.5 it decreased to 0.974 $\pm$ 0.109. In this regard, it should be noticed that a purely visual examination of the radio records with the interferometric responses of Cas A and Cyg A distinctly demonstrates a more intensive signal from Cyg A than one from Cas A. Let us recall that results of many different measurements for the flux-density ratio of Cas A and Cyg A at 38 MHz from 1954 to 2019, that were published in literature, was depicted in Figure 5 by \citet{Stanislavsky2023}. This picture clearly shows the decrease in Cas A flux density over time.

\section{Secular decline of Cas A around the peak of radio emission}
The most important and puzzling feature of Cas A is a secular decline of its radio flux in time. Considering the model of adiabatic expansion by explosion, \citet{Shklovskii1960} first predicted a secular decrease in the flux density of Cas A of about 2\% yr$^{-1}$. Using the absolute measurements of the Cas A flux density made around 1965 to an accuracy of $\approx$ 2\% over a frequency range from 22 MHz to 31 GHz, \citet{Baars1977} described the frequency dependence of Cas A fading rate for the 1965 epoch as $0.97 (\pm 0.04) - 0.30(\pm 0.04)\log(f/{\rm GHz})$ \% yr$^{-1}$. \citet{Erickson1975} were among the first to verify this formula in observations. Their Cas A flux-density measurements at 38 MHz in 1974 have shown that Cas A was stronger than the value predicted by \citet{Baars1977}. Subsequent studies of \citet{Read1977a,Read1977b}, \citet{Walczowski1985}, \citet{Rees1990}, \citet{Vinyajkin1997} and others at 38 MHz confirmed that the fading rate has changed from 1.9 $\pm$ 0.5 \% yr$^{-1}$  in the 1960s to 0.55 $\pm$ 0.14 \% yr$^{-1}$ in 2019 \citep{Stanislavsky2023}. Similar fading rates have been established at higher frequencies, and their values were collected by \citet{Reichart2000}. For frequencies below 38 MHz, much less trusted information on the fading rate is available due to intense RFI. The flux-density values of Cas A in the 1966 epoch at the frequencies of 12.6 MHz, 14.7 MHz, 16.7 MHz, 20 MHz and 25 MHz were obtained by \citet{Braude1969} using the Ukrainian T-shape Radiotelescope, with the first modification (UTR-1). Follow-up observations were continued with the Ukrainian T-shape Radiotelescope, second modification (UTR-2), in the 1970s and in the 1980s. Their extensive collection was presented by \citet{Bovkoon2010}. However, the frequency grid was too sparse, and it is in that frequency range where the maximum of the Cas A spectrum is located. The GURT measurements made it possible to overcome this disadvantage \citep{Bubnov2014,Stanislavsky2023}. This allows us to study the position of the spectral maximum with high accuracy.

\begin{figure}
\centerline{\includegraphics[width=\columnwidth]{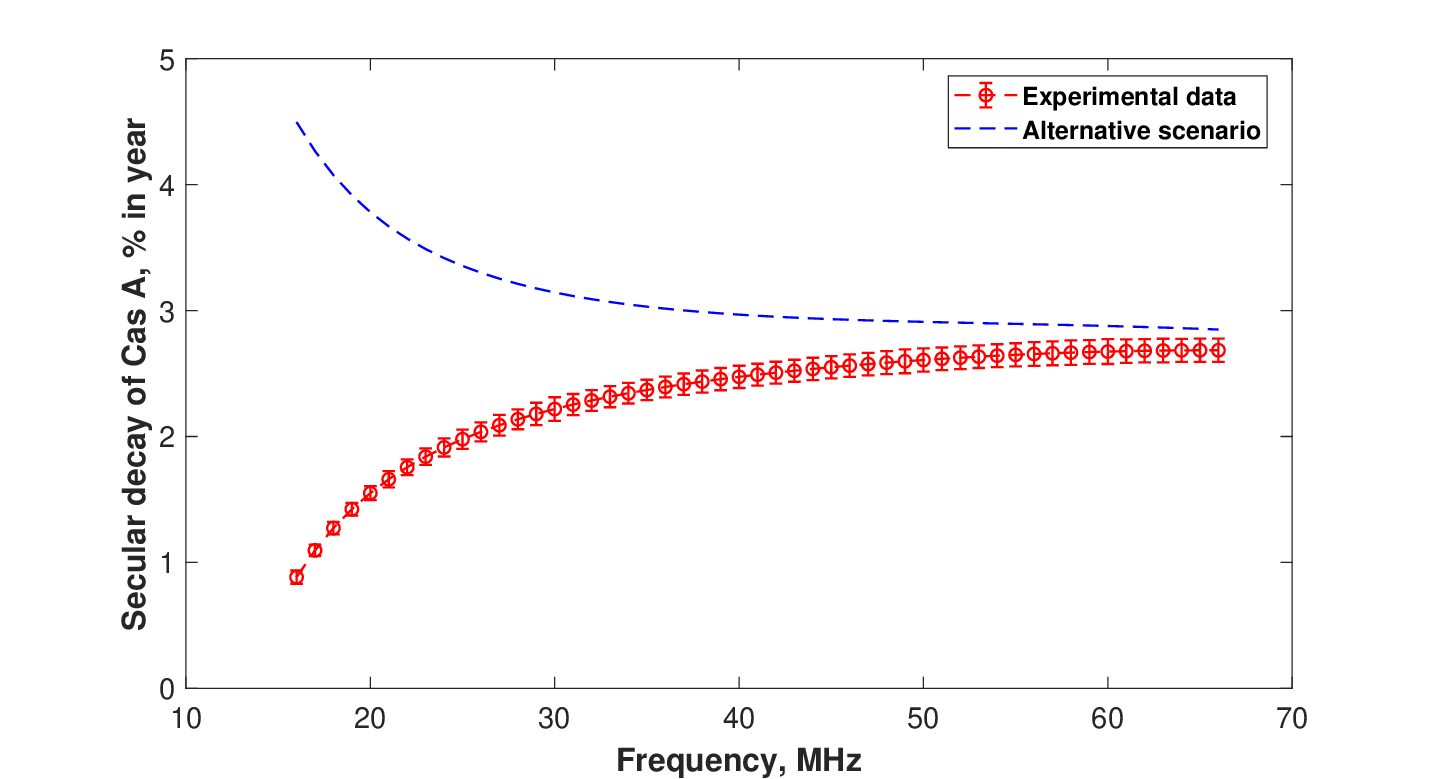}}
\caption{\label{fig8_rate}
Annual decrease in the Cas A flux density over the period 2019--2023.5 by frequency from radio observations. The blue dashed line shows how the annual evolution of the Cas A flux density would change in the period 2019--2023.5 if the emission measure $EM_{\rm ISM}$ increased by the same amount, rather than decreased. Then the spectral maximum would be located at $\sim$23 MHz.}
\end{figure}

The rate of decrease in the flux density of Cas A for the period {\it Year1}--{\it Year2} can be calculated by using the following expression
\begin{equation}
S_{\nu,{\it Year2}}=S_{\nu,{\it Year1}}\left(1-\frac{p}{100}\right)^n\,,\label{eq10}
\end{equation}
where $S_{\nu,{\it Year1}}$ and $S_{\nu,{\it Year2}}$ are the flux densities in {\it Year1} and {\it Year2}, respectively. Here, $p$ is the annual decrease in flux density in terms of percent, and $n$ denotes the observation period in years. Recent observations of 2019 and 2023.5 (\citealp{Stanislavsky2023} and in this paper) have highly overlapping frequency ranges and similar measurement accuracy. In this case the secular decay of Cas A around the turnover of radio emission at frequencies of 16--66 MHz is shown in Figure~\ref{fig8_rate}. The results of calculations according to Eq. (\ref{eq10}) noticeably depends on whether the maximum of the spectrum is moving along the frequency axis or standing still, and also in which direction it moves, towards high or low frequencies. The main behavior of the annual decrease in Figure~\ref{fig8_rate} was due to a decrease in $EM_{\rm ISM}$. If it had grown by the same number of values as it had decreased, then the situation would have looked the other way around. We simulated this scenario and showed it as a dashed line on the same figure. We note that in this case the secular decay would be greater.

\begin{table*}
\caption{Radio flux densities (in Jy) with Cas A according to measurements in the different epochs. These data were obtained from \citet{Bovkoon2010} and \citet{Bubnov2014}.}%
\label{tab2}
\centering
\small
\resizebox{\linewidth}{!}{
\begin{tabular}{p{2.5cm}p{2.5cm}p{2.5cm}p{2.5cm}}
\hline\hline
Epoch of & \multicolumn{3} { | c }{Frequency, MHz}\\
\cline{2-4}
observations & \multicolumn{1} { | l }{16.7} & 20 & 25\\
\hline
1959.5 & 53876 $\pm$ 2806 & 56775 $\pm$ 2972 & 54839 $\pm$ 2359\\
1962.7 & 51912 $\pm$ 2525 & 56180 $\pm$ 2972 & 48942 $\pm$ 2064\\
1967.9 & 56401 $\pm$ 3928 & 54397 $\pm$ 3270 & 48647 $\pm$ 2359\\
1971 & 56121 $\pm$ 4490 & 54397 $\pm$ 2675 & 48352 $\pm$ 1769\\
1974.5 & 56121 $\pm$ 3928 & 57369 $\pm$ 2972 & 51890 $\pm$ 2653\\
1978 & 51912 $\pm$ 4490 & 54991 $\pm$ 2675 & 50711 $\pm$ 2653\\
1986 & 56121 $\pm$ 4490 & 55586 $\pm$ 2675 & 48057 $\pm$ 1769\\
\hline
\hline
Epoch of & \multicolumn{3} { | c }{Frequency, MHz}\\
\cline{2-4}
observations & \multicolumn{1} { | l }{12.6} & 14.7 & 16.7\\
\hline
2009.4 & 37452 $\pm$ 1064 & 41002 $\pm$ 1025 & 39004 $\pm$ 842\\
\hline
\hline
Epoch of & \multicolumn{3} { | c }{Frequency, MHz}\\
\cline{2-4}
observations & \multicolumn{1} { | l }{35} & 45 & 55\\
\hline
2014 & 28975 $\pm$ 659 & 23616 $\pm$ 463 & 19535 $\pm$ 308\\
\hline
\end{tabular}
}
\end{table*}

\section{Discussion}

\subsection{Evolution of the Cas A spectral turnover}

Although the integrated radio spectrum of Cas A has been studied for a long time \citep{Baars1977,Kassim1989,Kellermann2009}, the evolution of its maximum at low frequencies remains an open question. First of all, it should be noted that its position in frequency was often only approximately known. This is due to the fact that direct measurements of this turnover frequency are rather complicated. In the era before digital receivers and low-frequency broadband antennas, most observations of Cas A were made at single frequencies in a narrow band. To find the secular change in the radio emission flux of Cas A at frequencies $\geq$ 38 MHz, this was quite enough. However, to analyze changes in the spectral maximum, such a frequency grid was too sparse, especially in the frequency measurements of the peak position. The most complete collection of measurement data for this radio source at the frequencies 12.6, 14.7, 16.7, 20 and 25 MHz from 1959.5 to 2009.4 can be found at \citet{Bovkoon2010}. It is in this frequency range that the maximum of the Cas A spectrum is observed. Table 1 of this work contains results of both widely known articles and unpublished measurements with UTR-2. However, they have different numbers of measurement frequencies. We have collected the most suitable ones in Table~\ref{tab2}. 

\begin{figure}
\centerline{\includegraphics[width=\columnwidth]{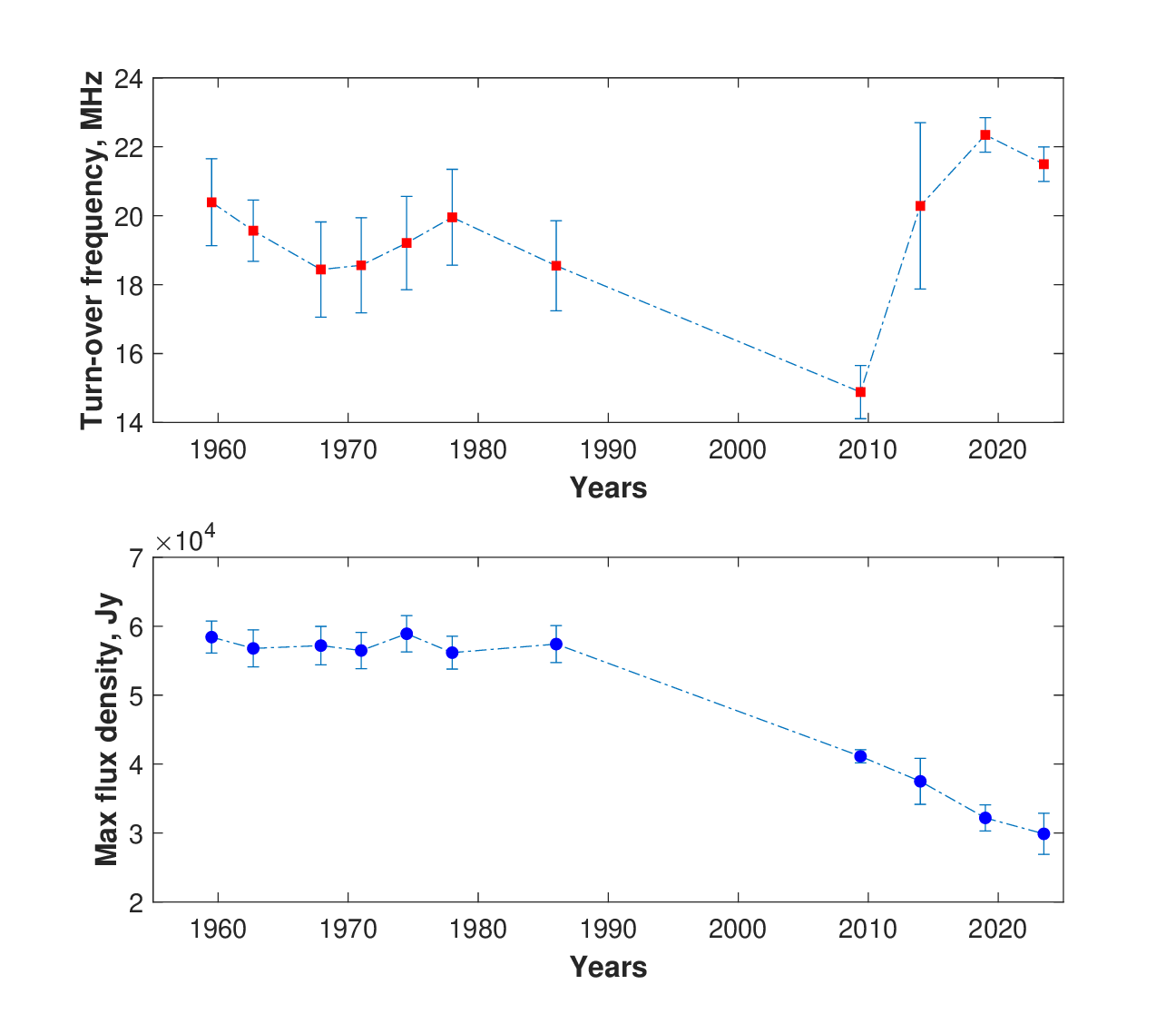}}
\caption{\label{fig9_max}
Turnover flux density and its frequency from different measurements (Table 1 of \citealp{Bovkoon2010}), the most of which was made at single frequencies, and the full integrated flux density is restored with Eq.(\ref{eq11}). The records of 2019 and 2023.5 give the measured values directly.}
\end{figure}

From those measurements, we can estimate the frequency position of the spectral maximum using modeling. The accuracy of these measurements is lower than that of observations from 2019 and 2023.5, and we can also notice that the frequency grid is sparse. Therefore, the use of Eqs.(\ref{eq5})--(\ref{eq8}) is hardly justified, if at all possible. However, it can be simplified by considering the expression of $J_\gamma(z)$ when $z\gg 1$. Then $J_\gamma(z)\sim z^{0.5-\gamma/2}$. Next, since in the frequency range under consideration, the optical depth $\tau_{\nu,\rm int}$ in the unshocked ejecta is noticeably more than one, the term $S_{\nu,\rm back}e^{-\tau_{\nu,\rm int}}$ can be neglected. According to \citep{Mezger1967} and \citep{Beckert2000}, the Gaunt factor $g_{\rm ff}$ is conveniently approximated by $\sim\nu^{-0.1}T^{0.15}$ with an accuracy of no more than 25\%. Consequently, we come to a simplified model, characterized only by three parameters. Its final form is very similar to the well-known expression (see, for example, \citealp{Bazelyan1965,Vinyaikin1987,Kassim1989}), namely
\begin{equation}
S_\nu\approx S_{0}\left(\nu/\nu_c\right)^\alpha\,\exp\left[-\tau_{\nu_c}(\nu/\nu_c)^{-2.1}\right]\,,\label{eq11}
\end{equation}
where $S_\nu$ is the integrated flux density at frequency $\nu$, $\nu_c$ denotes a reference frequency, and $S_{0}$ is constant that can be related to the known flux density at the spectrum maximum (or turnover), namely, $S_0 = S_{\nu_{peak}}\,\exp(|\alpha|/2.1)$. Here $\alpha = 0.5-\gamma/2$. Using the data of Table~{\ref{tab2}, we find the turnover frequencies of spectra that are assumed to have occurred in different years. We note also that ``turnover frequency'' is where the optical depth is equal to $|\alpha|/2.1$. The results of this analysis are presented in Figure~\ref{fig9_max}. Over the entire period of observations, the maximum of the spectrum decreased by approximately two times, and the turnover frequency  fluctuates in the range between 15.2 MHz and 22.34 MHz. This may indicate that the spectral maximum of Cas A is not fixed in time and frequency, and this frequency evolution is nonmonotonic.

\begin{figure}
\centerline{\includegraphics[width=\columnwidth]{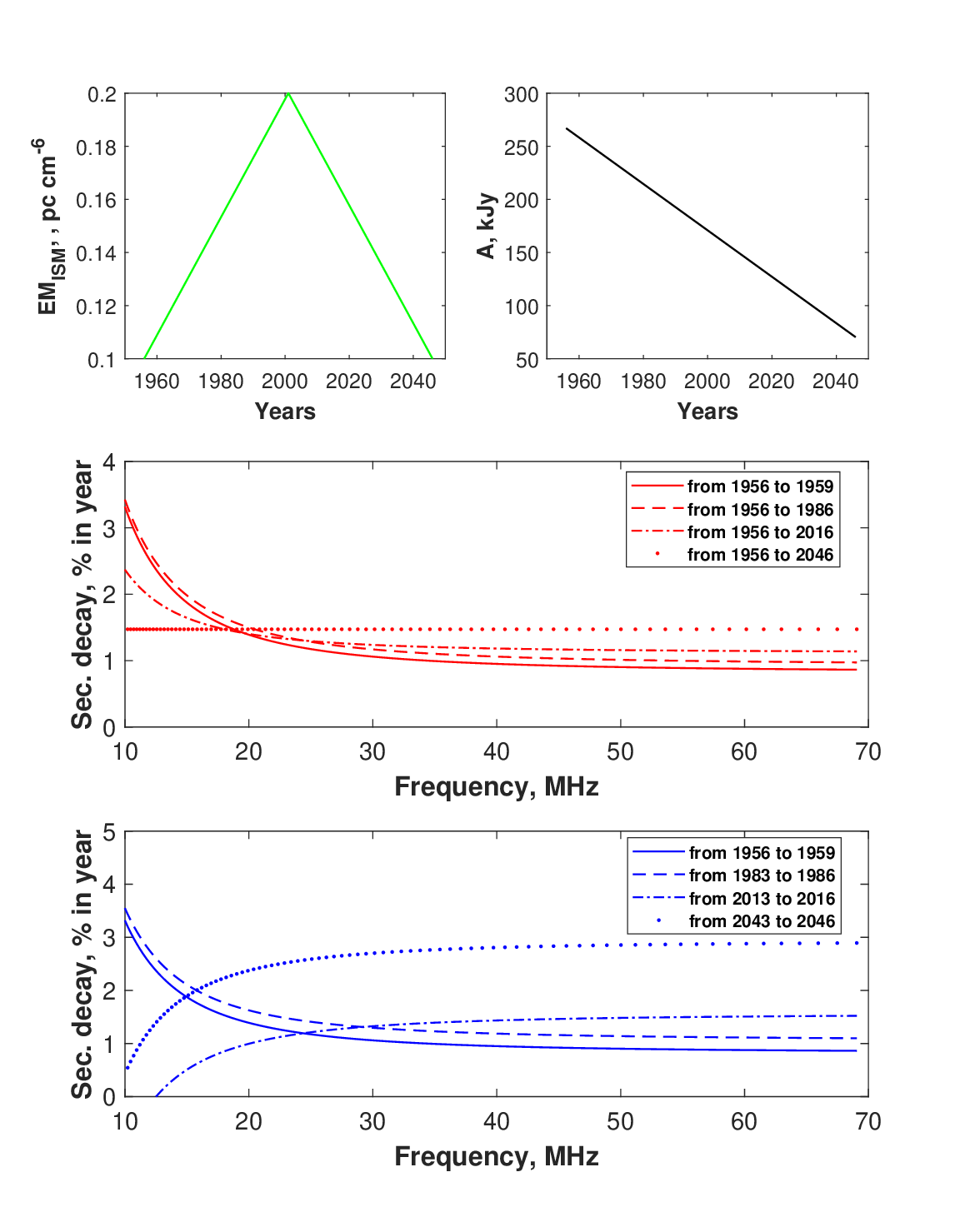}}
\caption{\label{fig9_model}
Modeling of flux-density changes in Cas A over the period 1956--2046 at 10--70 MHz. Secular decline is determined from Eq.(\ref{eq10}). Evolution of the Cas A spectrum is driven only by two parameters: $EM_{\rm ISM}$ and $A$ characteristic for the ``bare'' radio source (see Subsection~\ref{model}).}
\end{figure}

\subsection{Modeling the secular decline at low frequencies}
If we compare the spectral flux-density measurements of Cas A in 2023.5 with the data of \citet{Braude1969} in the epoch of 1966 (as in Table 2 of \citealp{Stanislavsky2023}), then the secular decline of the flux density emitted by this SNR in the frequency grid leads to values similar to ones of 2019. This can be explained by the position of the spectral maximum at $\sim$18 MHz in 1966, close dates (2019 and 2023.5) and small changes in flux density between them. However, if we imagine that the downward trend in the ISM emission measure will continue, then the secular decline of the flux density for Cas A may demonstrate a different behavior in the future. This scenario has been modeled as follows. For the sake of simplicity we assume that most parameters are unchanged, i.\,e. $EM_{\rm int}$ = 37.4 pc cm$^{-6}$, $T_{\rm int}$ = 100 K, $Z_{\rm int}$ = 2.5, $f_a$ = 0.8, $T_{\rm ISM}$ = 20 K, and $Z_{\rm ISM}$ = 0.5 as well as $\gamma\approx2.52$ and $\nu_1\approx6.15$~MHz (which are close to the real values). The ISM emission measure evolves over time between 0.1 and 0.2 pc cm$^{-6}$. We let it increase linearly from 1956 to 2001, and then it returns to its initial value in 2046. The decrease in the flux density of Cas A is achieved due to a linear decrease in $A$ from 267 kJy to 70 kJy (see Subsection~\ref{model}). In this case the frequency position of the Cas A spectral maximum reaches 23.94 MHz in 2001 and returns to its original value of 17.75 MHz in 2046. Then the peak of the modeled spectrum monotonically decreases from 67340 Jy in 1956 to 17720 Jy in 2046, which approximately corresponds to the expected values. Using Eq.(\ref{eq10}), we find the secular decline of the Cas A flux density by frequency in two ways: from 1956 to subsequent years until 2046 as well as on the grid of years between nearest counts with a step of three years (Figure~\ref{fig9_model}). From these simulations, it can be seen that the secular decline from 1956 to subsequent years is especially large at the lowest frequencies at first. However, thanks to the symmetry of evolution in $EM_{\rm ISM}$, the secular decline becomes the same at 10--70 MHz. On the grid of years we first observe that the magnitude of the secular decline between nearest counts with a step of three years is higher at the lowest frequencies than at the highest. Yet, after 2001, when $EM_{\rm ISM}$ starts decreasing, the secular decline behaves exactly the opposite way. Following Figure~\ref{fig9_max}, real changes of $EM_{\rm ISM}$ over time may be more complex, repeatedly increasing and decreasing in magnitude. As a results, the secular decline of Cas A shows the course of $EM_{\rm ISM}$.

All these results indicate that free-free absorption by the ISM ionized gas along the line of sight to the source influences on the secular decline of Cas A. One of the possible reasons for such changes may be the interaction between the SNR and its environment. As has been remarked by \citet{Castelletti2021}, interactions between SNRs and their environs are readily detectable through thermal absorption using low frequency observations of SNRs with high sensitivity and resolution. The NenuFAR together with URAN--2 just make reliable measurements below 100 MHz. Their low-frequency measurements of SNRs can have a major impact on this field. There are many cases in which finding the low frequency turnover for any SNR requires more sophisticated measurements than a single low frequency measurement. Thus, our observations firstly have been detected this interaction between Cas A and its surrounding gas as an absorber from its radio spectrum at low frequencies.

\section{Conclusions}
We have established the integrated radio continuum spectrum of Cas A at 8--66 MHz in the epoch of 2023.5. The measurements were implemented with two different radio arrays each of which was a two-element correlation interferometer. To provide the measurements at 21--66 MHz the NenuFAR array was applied, and two equal halves of the URAN--2 array carried out interferometric observations at 8--40 MHz. The spectrum of Cas A was obtained by connecting these curves. The radio galaxy Cyg A served as a reference radio source whose known spectrum was assumed to be constant. Fitting the spectral model of Cas A with free-free absorption effects from the ISM and interior unshocked ejecta, the parameter values ($EM$, $T$, $Z$, and $f_a$) of internal and external absorption were found. Comparing our measurements with published results for the flux density of Cas A in a wide frequency band, the secular decline was detected and matches the expected. We summarize our results as follows:

\begin{enumerate}
      \item The measured spectrum of Cas A with NenuFAR and URAN--2 interferometric observations is consistent with the theoretical one within the measurement errors. The spectrum of Cas A is determined by not only the synchrotron emission mechanism, but by thermal absorption inside the SNR and in the ISM.
			\item In the epoch of 2023.5 the peak of the Cas A flux-density spectrum is 29880 $\pm$ 2980~Jy at 21.5 $\pm$ 0.2 MHz. This frequency shift of the peak in comparison with the GURT observations in 2019 \citep{Stanislavsky2023} indicates that Cas A, expanding in space, interacts with CSM having a lower density than before. 
			\item From these measurements the area, associated with unshocked, photoionized ejecta, internal to the reverse shock, has a temperature $T\approx 100$ K, an average ionization state of $Z\approx 2.5$ and an average emission measure of $EM\approx 37.4 \, \rm{pc} \, \rm{cm}^{-6}$. 
			\item We find the ISM emission measure $EM\approx 0.15 \, \rm{pc} \, \rm{cm}^{-6}$ along the line of sight to Cas A for a 20 K ISM. Observations of \citet{Stanislavsky2023} established that 86\% of the emission emerging from the region projected against the unshocked ejecta came from the foreground side of Cas A, and \citet{Arias2018} obtained 78\%. New measurements of 2023.5 gives a value between them, 80\%, but it is closer to the result of \citet{Arias2018}.
			\item The measured free-free absorption parameters, mentioned by \citet{Arias2018} and represented by \citet{Stanislavsky2023}, are very close to the experimental estimates for their values from the recent observations with NenuFAR and URAN--2.
			\item In the frequency range of 8--66 MHz, the flux density of Cas A decreases faster at higher frequencies than at lower frequencies in comparison of observations between the epoch of 2019 and 2023.5. This is explained by the decrease in the ISM emission measure. Consequently, the maximum of the Cas A spectrum is shifted to lower frequencies. If in the epoch of 2019 the spectral index was about 0.59 $\pm$ 0.02 in the frequency range of 38--70 MHz, then in 2023.5 its value is 0.64 $\pm$ 0.07, i.e. these values are sufficiently similar. 
			\item It should be noted that measured values of the free-free absorption parameters towards Cas A are consistent with other recently published results \citep[and references therein]{Arias2018,Stanislavsky2023}. Our results are useful because they support previous results in a completely independent way with a completely different instrument. We applied a new combination of instrumental tools and extend the frequency range of observations, which was not available in previous studies with GURT.
\end{enumerate} 

\begin{acknowledgements}
LS acknowledges the support of young researchers due to Contract No. 0123U103014 from the National Academy of Sciences of Ukraine. AS and IB thank the National Academy of Sciences of Ukraine for the partial support under Contract No. 0123U102426. Personally, we are grateful to O. Ul'yanov for his help in the organization of observations with URAN--2 and useful discussions. We also thank the anonymous referee for very valuable comments that contributed to improve the quality of the manuscript.

This paper is based on data obtained using the NenuFAR radio-telescope. The development of NenuFAR has been supported by personnel and funding from: Station de Radioastronomie de Nan\c{c}ay, CNRS-INSU, Observatoire de Paris-PSL, Universit\'e d’Orl\'eans, Observatoire des Sciences de l’Univers en r\'egion Centre, R\'egion Centre-Val de Loire, DIM-ACAV and DIM-ACAV+ of R\'egion Ile de France, Agence Nationale de la Recherche.

The Nan\c{c}ay Radio Observatory is operated by the Paris Observatory, associated with the French Centre National de la Recherche Scientifique (CNRS).
 
\end{acknowledgements}



\bibliographystyle{aa} 

\end{document}